\documentclass[prb,preprint]{revtex4-1}

\usepackage{amsmath}  
\usepackage{amsfonts} 
\usepackage{graphicx} 

\begin{document}


\title{Diffusion as a First Model of Spread of Viral Infection}

\author{Paulo H. Acioli}
\email{p-acioli@neiu.edu} 
\affiliation{Department of Physics and Astronomy, Northeastern Illinois University, Chicago, IL 60625}


\date{\today}

\begin{abstract}
The appearance of the coronavirus (COVID-19) in late 2019 has dominated the news in the last few months as it developed into a pandemic. In many mathematics and physics classrooms, instructors are using the time series of the number of cases to show exponential growth of the infection. In this manuscript we propose a simple diffusion process as the mode of spreading infections. This model is less sophisticated than other models in the literature, but it can capture the exponential growth and it can explain it in terms of mobility (diffusion constant), population density, and probability of transmission. Students can change the parameters and determine the growth rate and predict the total number of cases as a function of time. Students are also given the opportunity to add other factors that are not considered in the simple diffusion model.
\end{abstract}

\maketitle 

\section{Introduction} 
The end of 2019 and beginning of 2020 have been dominated by the spread of the coronavirus (COVID-19). The disease started in the Hubei province in China in December 2019 and by early January 2020 started to spread. It grew very rapidly, triggering responses from the Chinese and other governments. On March 11, 2020 the World Health organization declared that COVID-19 was then characterized as a pandemic.\cite{pandemic} The situation triggered different responses from different governments. In The United States, many colleges took the initiative to start their own social distance programs, including sending all students home, extending spring breaks, and ultimately moving all classes to distance learning.\cite{response1, response2} Many cities and states followed up with shelter at home mandates. \cite{response3} The situation in many countries became alarming due to the exponential growth of new cases and deaths. \cite{JH,WHO}

\begin{figure}[h!]
\centering
\includegraphics[scale=1]{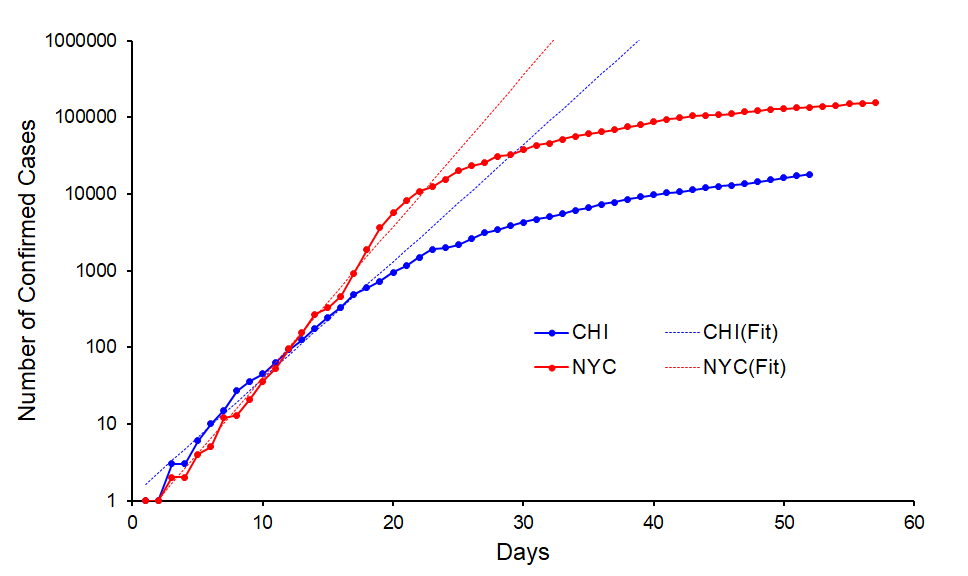}
\caption{COVID-19 cases in Chicago (CHI,blue)\cite{CHI-C19} and New York City\cite{NYC-C19} (NYC,red). The dashed lines are the best fit to the date after the initial spike in cases. The $x$-axis is the number of days from the date that the first case in each city was detected. }
\label{fig1}
\end{figure}

One of the consequences of this pandemic is that instructors at colleges and universities started to monitor and model the data, using this as a teachable moment for students and colleagues. The first step in modeling the data is to plot the number of cases as a function of time and show that it exhibits an exponential growth. For beginners it is important to introduce them to the logarithmic scale, where the plot becomes a straight line and students can extract the exponent and realize that the fit to the US data on March 20, 2020 shows that the number of cases doubled every 2.4 days. Fits to the initial data for Chicago and New York City show that New York has a higher growth rate than Chicago FIG. \ref{fig1}, and the total number of cases in Chicago is substantially smaller. One can speculate that this might be due to difference in population densities in these cities. After the first 20 days and after mitigation and social distance measures were imposed, it is clear that the rate of infection slowed down.

In this work we propose to look at some of the factors affecting the spread of viruses using a simple diffusion model in which each individual in a population is treated as a Brownian \cite{brownian} particle with diffusion constant $D$. Also added to this model is the incubation period of the virus and a probability of transmission of the virus if individuals are closer than a certain distance. This model is to be used as a project in a computational physics course and verify if it adequately predicts the exponential growth of the number of cases, as well as if the population density, mobility, and probability of transmission play roles on the percentage of the population that will be infected as a function of time. Students, will be asked to modify a code, analyze the output for different sets of parameters, and write a critical analysis of its predictive effectiveness.

\section{Computational Project Details}
In this section we propose a project to be implemented in computational physics courses to study the spread of infectious diseases as a simple diffusion of individuals. The benefits of this project is that its implementation is simple, but it can lead to a qualitative understanding on how diseases are spread and it can also allow the student to understand factors that can affect it. The main ingredients of the model are: individuals are considered particles that obey a Brownian diffusion process; each individual will have three possible states, healthy, sick (contagious), and cured; a healthy individual has a probability of getting infected if its distance to a sick individual is smaller than a certain threshold; the incubation and sickness periods are the same; once an individual gets cured, it cannot be infected or contagious again. VPython (or GlowScript)\cite{vpython} is the language of choice as it allows for a real time visualization of the infection spread and it will allow for fast simulations of small populations on a laptop. In the assignment, students will implement the code, analyze the data generated, critique the initial assumptions, and propose improvements for more realistic simulations. Below we describe the algorithm starting with the standard diffusion equation.

The diffusion equation in 2D
\begin{equation}
\frac{\partial{ f(x,y,t)}}{\partial{t}} = D\left[ \frac{\partial^2 f(x,y,t)}{\partial{x^2}}+\frac{\partial^2 f(x,y,t)}{\partial{y^2}}\right] \label{eq1}
\end{equation}
is used to study many phenomena\cite{ajp1,ajp2,ajp3,ajp4,ajp5,ajp6,ajp7,ajp8,ajp9,ajp10} from diffusion inside the nucleus to population dynamics to solving the Shr\"odinger equation. The function $f$ will have different meanings depending on the application, from temperature to density. In this work we will use it for the diffusion of individuals, treated as particles, over closed boundaries subject to contamination of a viral infection. The normalized solution to Eq. (\ref{eq1}) for a single particle is given by
\begin{equation}
f(x,y,t)=f_0\frac{1}{\sqrt{4\pi Dt}}e^{-\frac{x^2+y^2}{4Dt}}. \label{eq2}
\end{equation}
Therefore one can simulate the diffusion of a particle from its previous position by generating a Gaussian distribution of zero mean and variance $\sqrt{2Dt}$. For a system on $N$ non-interacting particles with the same diffusion constant we use Eq. (\ref{eq2}) for each particle at each simulation time step.

The next ingredients in the simulation will be the population density $\rho$, the number of habitants (particles in the simulation cell) $Npop$, the diffusion constant $D$, the number of simulation steps $Nstep$, the time step $dt$, the incubation period ($t_{inc}$), the transmission radius, and the probability of transmission from an infected to a healthy individual ($prob$). All of these variables are set at the beginning of the simulation. We preset that the total simulation time consists of 90 days and that each time step is 0.01 days, therefore each simulation takes 9000 steps. The algorithm is described below:
\begin{enumerate}
\item Input $Npop$, $Nstep$, $D$, $dt$, $t_{inc}$, $prob$, $r_{transm}$
\item Calculate the size of the square cell as $L=\sqrt(Npop/\rho))$
\item Initialize the initial population
\item Choose a fraction of the initial population to be infected, and set timer for the sickness ($t_{sick}$)
\item Loop over $Nstep$
\item Move all individuals according to the Gaussian distribution (Eq. (\ref{eq2})).
\item Compute the distance between each healthy and infected individuals
    \subitem If the distance is less than $r_{transm}$, the healthy individual becomes sick with probability $prob$
\item Subtract the sickness timer by $dt$
\item If $t_{sick}<0$ the sick individual gets cured

\end{enumerate}

In order to generate more accurate statistics we suggest the students run simulations with the same initial parameters multiple times, between 20 and 100, depending on the size of the system and the speed of the student's computer. We suggest that the smallest population contains 100 individuals, as even one sick individual corresponds to an initial infected population of 1\%. Depending on how much time the students have to complete the analysis of the project, they can use a population of 1000, being aware that each individual 9000 steps simulation can take up to a 30 minutes. They can speed up the process by using larger time steps, however, they must test if the results of an individual simulation with the same diffusion constant and different time steps lead to similar outcomes.

\section{An Example of a Project}
In this section we provide an example of a project that students could pursue and also validate the method by comparing the results of the simulation with the data presented in FIG. \ref{fig1}. In this project, students will study the differences of the infection proliferation in New Your City (NYC) and in Chicago. New York City was chosen because it is experiencing a very rapid growth on the number of cases and preventive measures such as shelter at home were taken by the city and state governments at very early stages of the infection. Chicago was chosen as a local connection to Northeastern Illinois University (NEIU) students, and is also experiencing an exponential growth in the number of cases and has also been affected by shelter at home mandates. We must be mindful that the current model will not be able to study the effects of the measures that the government are taking, but it might be able to justify their need.

The population density of New York city is 10,194 people/km$^2$,\cite{nycp} while Chicago's is  4,665 people/km$^2$.\cite{chip} It is natural to offer the students the hypothesis that if all other variables are the same, the spread of infections in New York City will be faster than in Chicago and that the number of cases will be much larger for the same period of time after the first case. With this hypothesis alone, students should be able to generate a enough data for the project. They can also discuss if the mobility (diffusion constant) should be the same in both cases and study the effect of mobility in the spread of the disease. In the proposed model the period of incubation is the same as the period of sickness, students can be offered the option to modify this assumption. The model also assumes that once cured, an individual will attain immunity, and will not be able to spread the disease. They can discuss modifications to the model to incorporate relapse. It is clear that with this very simple model and this very limited two-city project, they can perform a very thorough study that can give us some qualitative understanding of infectious disease spread.

In order to validate our model we will try to recreate the same initial exponential growth as experience by Chicago and New York City. The exponential fit to the data in FIG. \ref{fig1} yields exponents of 0.352 and 0.455 for CHI and NYC, respectively. Although we envision that we can see such growth in a 100 individual population, this growth is unattainable for the same time frame as the real data, as 100\% of the population would be infected in days 13 and 11 if we start with one individual infected in day 0. Deviations from the exponential behavior will be seen even earlier as there is not enough susceptible population available to be infected. In order to get similar exponential growth we start with simulations of populations of 100 individuals and 1 sick individual chosen at random. The initial diffusion constant was chosen as $D = 100$m$^2$/day with a time step of $dt=0.01$day. Therefore the variance of the gaussian distribution is $\sqrt{2}$. In this case the number of simulation steps is 9000 for each individual simulation is  27s in a laptop with Intel(R) Core(TM) i5-8250U CPU @ 1.60GHz. Therefore 50 simulations will take about 22.5 minutes of computation. We assume that the radius of contamination is 2m, and that the probability of contamination is 20\% per time step. The incubation period is taken as 14 days.  Below are some of the results of these simulations.

\begin{figure}[h!]
\centering
\includegraphics[scale=1]{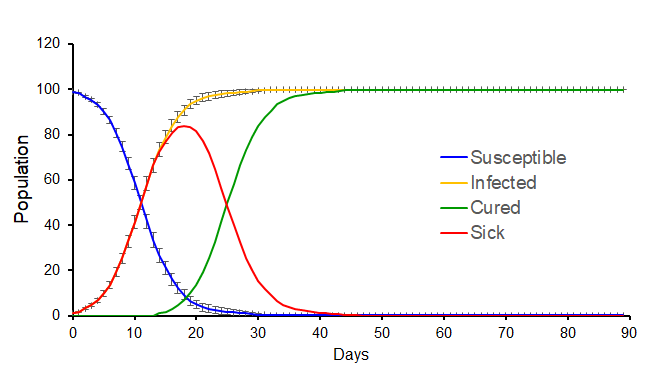}
\includegraphics[scale=1]{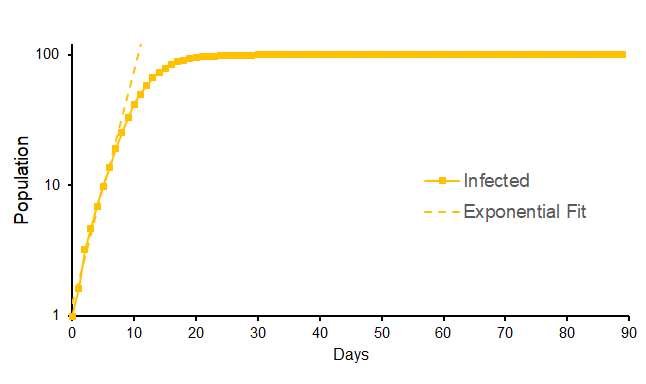}
\caption{a) Average of 50 90-day simulations of the spread of a virus in a population of 100 individuals in a square cell with the population density of NYC ($\rho=0.012$ people/m$^2$). $D = 100$m$^2$/day, $prob=0.2$, $dt=0.01$day. b) Graph of the number of infected individuals (full line) in a logarithmic scale and the exponential fit for the initial 10 days of the simulation (dashed line).}
\label{fig2}
\end{figure}

In FIG. \ref{fig2}a we present the results of the simulation for New York City with the set of parameters in the previous paragraph. As one can see, under this assumption 50\% of the population is sick on day 11 and 99\% of the population will be infected after 28 days. One can see that after day 18 the number of sick people starts to decline. However, if the death rate is similar to what has been observed on the COVID-19 pandemic, about 5.5\% of the population of NYC would perish, and the numbers would be even worse, since no major city in the world would be able to have hospital beds for 83.8\% of its population at the peak of the infection. In the figure we also observe that the rate of cure follows the number of infected individuals with a lag time of 14 days, which is the incubation/sickness period. An exponential fit, shown in FIG. \ref{fig2}b, yields an exponent of 0.41 which is a bit lower than what was observed in the real data. Students will need to search for the set of parameters that is able to fully match the initial growth rate data that they choose to model. 

\begin{figure}[h!]
\centering
\includegraphics[scale=1]{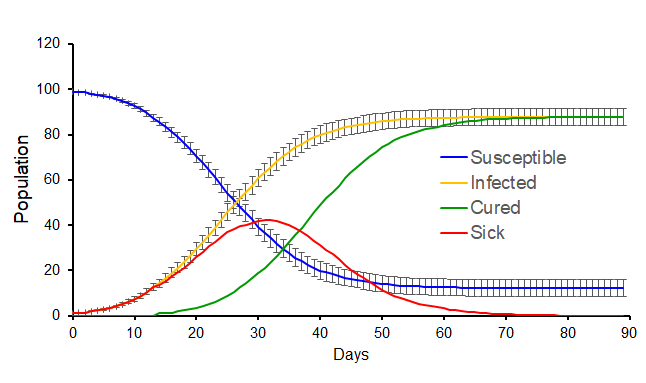}
\includegraphics[scale=1]{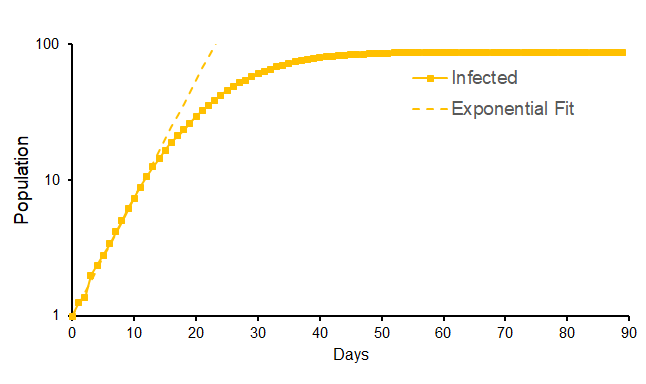}
\caption{Average of 50 90-day simulations of the spread of a virus in a population of 100 individuals in a square cell with the population density of the city of Chicago ($\rho=0.0047$ people/m$^2$). $D = 100$m$^2$/day, $prob=0.2$, $dt=0.01$day.b) Graph of the number of infected individuals (full line) in a logarithmic scale and the exponential fit for the initial 10 days of the simulation (dashed line).}
\label{fig3}
\end{figure}

In FIG. \ref{fig3}a we report the results for the city of Chicago. Most of the parameters are the same, with the exception of the density that will be changed to $\rho=0.0047$ people/m$^2$. For this set of parameters 50 \% of the population will be infected after 26 days reaching a maximum of 87.8\% after day 86.  The peak of the number of sick people is 42.3 \% and it happens about 32 days after the first 1\% of the population is infected. In FIG. \ref{fig3}b we present a logarithmic scale plot of the total number of infections together with and exponential fit for the initial stages of the infection. Similarly to what happened in the simulation for NYC it is clear that the number of infections deviates from the exponential growth after 13 days. The exponent in the fit is 0.2 which is also smaller than the value for the data for Chicago. A closer fit is likely to take place by increasing the probability of infection.

These two examples do support our initial hypothesis that under the same conditions, one would expect a slower growth rate in a less dense population. This leads to the conclusion that keeping all parameters the same, the percentage of the population infected is correlated to the population density. In order to determine how mobility affects the rate of infection, one would change the diffusion constant and repeat the simulations. The expectation is that the lower the mobility, the lower the total infection rate. In addition, even with the limitations of a small population simulation we were able to find growth rates of infections that are similar to what was observed for the cities of Chicago and New York. Thus validating this approach to simulate the spread of infectious diseases. 

\begin{figure}[h!]
\centering
\includegraphics[scale=1]{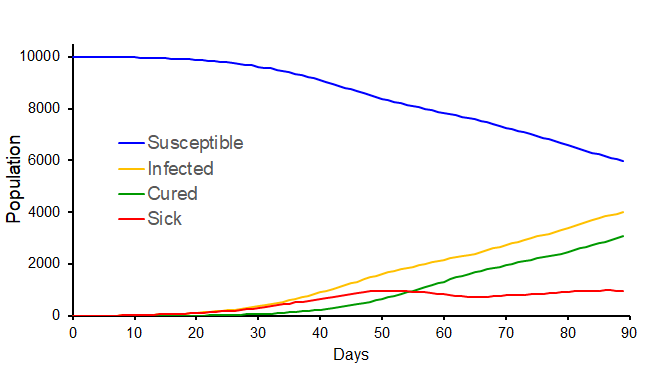}
\includegraphics[scale=1]{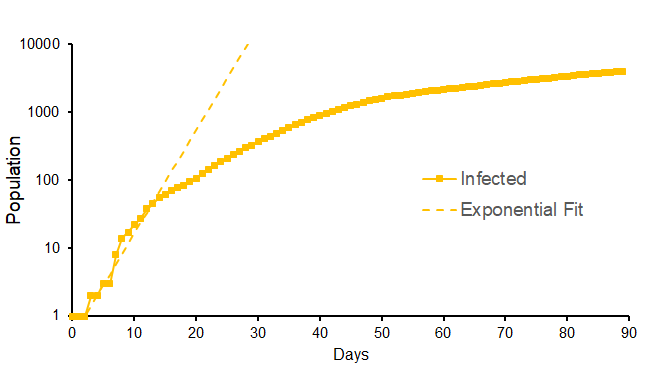}
\caption{One 90-day simulations of the spread of a virus in a population of 10000 individuals in a square cell with the population density of the city of Chicago ($\rho=0.0047$ people/m$^2$). $D = 100$m$^2$/day, $prob=0.3$, $dt=0.01$day. b) Graph of the number of infected individuals (full line) in a logarithmic scale and the exponential fit for the initial 15 days of the simulation (dashed line).}
\label{fig4}
\end{figure}

In order to show the limitations of small population simulations we run a simulation for the city of Chicago with a probability of infection of 30\% per time step but increased the number of individuals to 10000. The results are shown in FIG. \ref{fig4}a. With this set of parameters we obtain the same exponential growth as observed in FIG. \ref{fig1} for a period of 15 days.
Starting with 0.01 \% of the population infected at the end of 90 days about 40 \% of the population will have been infected. One can see that even a small diffusion constant of 100 m$^2$/day and a probability of transmission of 30 \%, the virus will spread to a large percentage of the population. It is clear that a diffusion model does a very good job of simulating the spread of the COVID-19 virus and could be used in a classroom setting, and with the modifications suggested below one can even use it for more realistic predictions when simulating larger populations. The main limitation will be access to better computational resources such as a parallelized code and a computer cluster.

\begin{figure}[h!]
\centering
\includegraphics[scale=0.55]{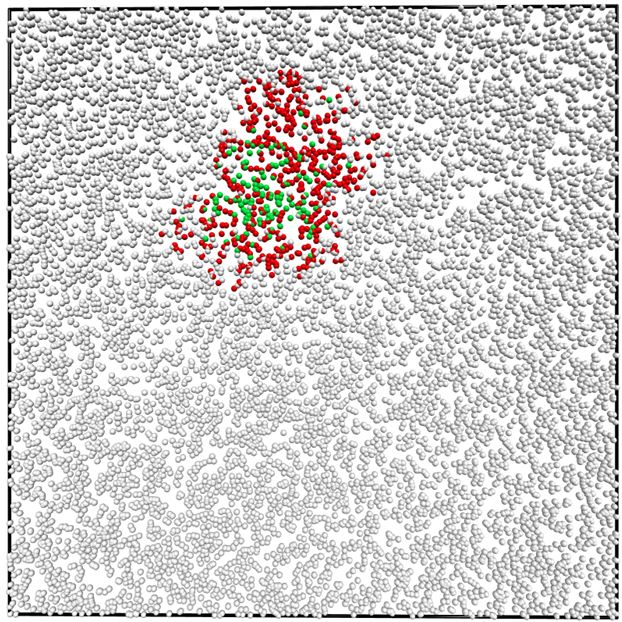}
\includegraphics[scale=0.55]{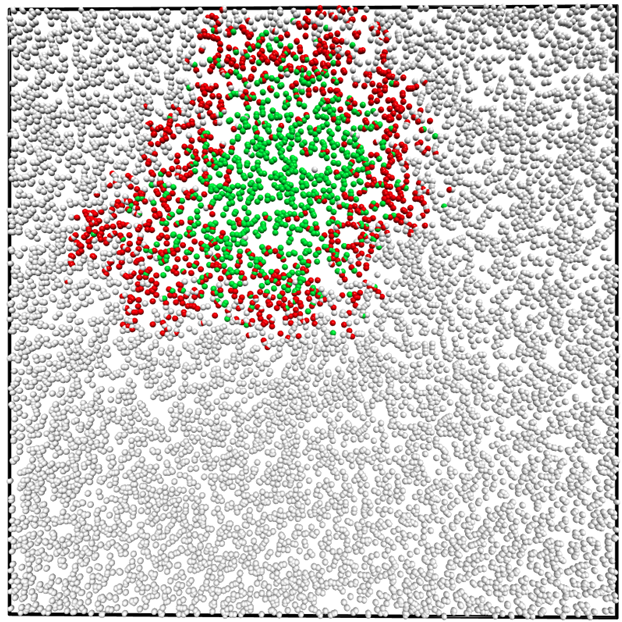}
\includegraphics[scale=0.55]{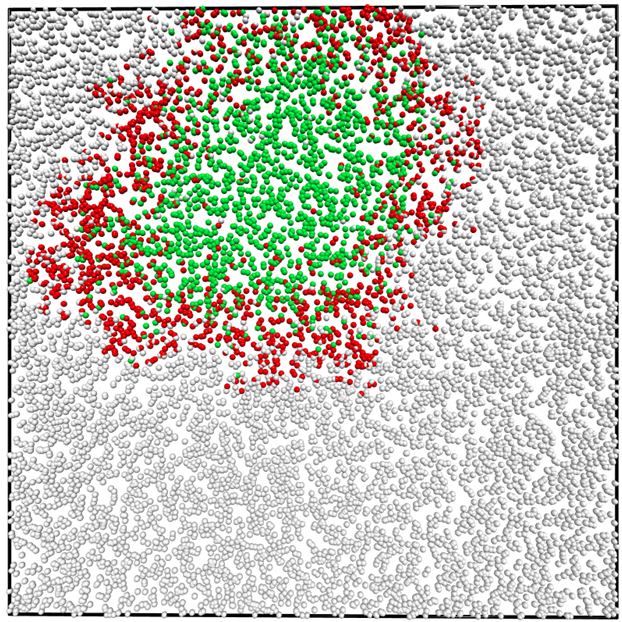}
\includegraphics[scale=0.55]{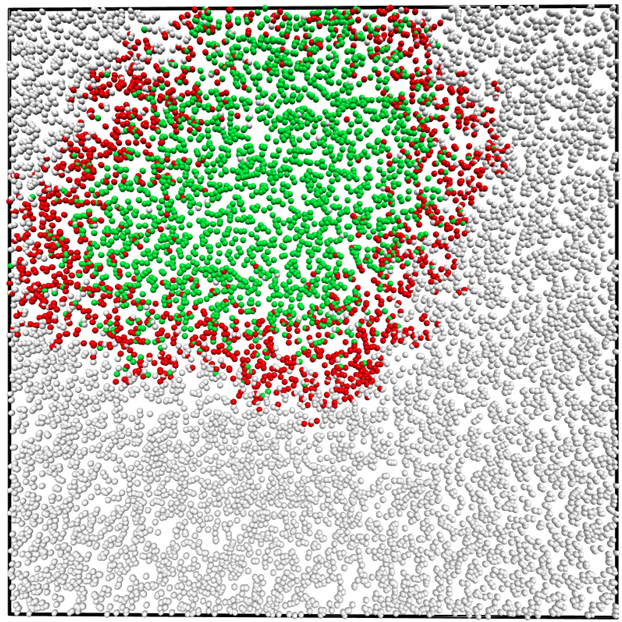}
\caption{Snapshots of one of the simulation 90-day simulations of the spread of a virus in a population of 10000 individuals in a square cell with the population density of CHI ($\rho=0.0047$ people/m$^2$). $D = 100$m$^2$/day, $prob=0.3$, $dt=0.01$day.}
\label{fig5}
\end{figure}

To finalize, we show snapshots of the simulation cell and its population in FIG. \ref{fig5}. White spheres represent the susceptible people, red represent sick, and green are those individuals that recovered. This window is very useful for a quick analysis of what is happening and can show how the infection is spread over time. We found it useful to change the radius of the individuals so that they are visible as the population increases, and the cell size on the screen remains the same. Because we are displaying the results for 10000 individuals it is interesting to see that the infection has a clear origin and the diffusion process spreads the disease outwards. In these simulations we chose closed boundaries and as a result there will be a limitation on the growth as there will be less susceptible neighbors to be infected when the disease reaches a boundary. To avoid this limitation one can use periodic boundary conditions.

The main limitation of this model is that it assumes the same mobility for all individuals and a constant density in the simulation cell, and no travel between cities. In addition, the average distance traveled by Brownian particles is proportional to the square of the simulation time. Thus, there is a limit on how far they will travel and how many people an individual can infect.  However, depending on the level of the students, instructors can use the same code and create sub-regions in the simulation cell that would have different population densities and subsets of the populations can be confined to these regions. The students could allow individuals to go from one region to another with a given probability. This would allow the possibility of an infected individual to move to a region that otherwise would have no infections. This modification would allow for the spread of the infection across borders. One could also include quarantine effects, by creating small cells where a single infected individual will be confined for a period of time and no other individuals are allowed in. Social distancing can be implemented by a small repulsive potential attributed to some individuals. These modifications will bring this simple model closer to more realistic simulations as those of references \cite{ndlib,infdiffusion}. To help the implementation of these projects, the original code can be downloaded from the author's website \cite{code}.

\section{Conclusion}
This project on itself is simple enough that it can be implemented and analyzed in a first course on computational physics. However, it is rich enough that can lead to a large amount of data that can be used to qualitatively and quantitatively analyze the spread of viral infection. In addition, it can help students understand the phenomenon of particle diffusion and Brownian motion. In the example presented, the simulation was limited to the same 20\% chance of infection, the same time step, the size of the population was fixed, and the length of the disease was the same as the incubation period. Under these conditions students will have a larger number of parameters to change and verify if they can reproduce data readily available in the news. In addition, they will have the opportunity to determine that the initial growth rate in the number of cases follows an exponential trend and notice that once it reaches a certain threshold if it will follow the expected logistic behavior.

In conclusion, we hope to have convinced the reader that a simple diffusion model can be used to qualitatively explain the spread of disease and even in some cases quantify it. It is our judgment that it will allow students to work on a problem that is directly affecting them and that these simulations will help them offer valuable insight on the issue.

\begin{acknowledgments}

I would like to thank Dr. Greg Anderson for sparking the discussion of using the COVID-19 pandemic as a teachable moment and Dr. Orin Harris for inspiring the idea of comparing the data for different populations.

\end{acknowledgments}

\end{document}